\documentclass[12pt,english]{iopart}
\usepackage{iopams}
\usepackage{cite}

\pdfoutput=1
\usepackage[pdftex]{hyperref}

\usepackage{graphicx}
\usepackage[normalem]{ulem}

\begin{document}

\title[An escape of vector matter-wave soliton from a parabolic trap]{An escape of vector matter-wave soliton from a parabolic trap}

\author{Yuliy V. Bludov$^1$, Monica A. Garc\'{i}a-\~{N}ustes$^2$}

\address{$^1$Department of Physics and Center of Physics, University of Minho, PT-4710-057, Braga, Portugal}

\address{$^2$QuantaLab, University of Minho, PT-4710-057, Braga, Portugal}

\address{$^3$Instituto de F\'isica - Pontificia Universidad Cat\'olica de Valpara\'iso, Avenida Brasil, Valpara\'iso, Casilla 2950, Chile.}

\ead{bludov@fisica.uminho.pt}

\date{}

\begin{abstract}
We show that a vector matter-wave soliton in Bose-Einstein condensate loaded into an optical lattice can escape from a trap formed by a parabolic potential, resembling a Hawking emission. The particle-antiparticle pair is emulated by a low-amplitude bright-bright soliton in two-component Bose-Einstein condensate with effective masses of opposite signs. It is shown that the parabolic potential leads to a spatial separation of BEC components. One component with chemical potential in a semi-infinite gap exerts periodical oscillations, while the other BEC component, with negative effective mass, escapes from the trap. The mechanism of atoms transfer from one BEC component to another by spatially periodic linear coupling term is discussed.



\end{abstract}

\submitto{\jpb}

\maketitle

\section{Introduction}

One of the prominent features of the Bose-Einstein condensate (BEC) subjected to the external potential is the possibility to use it as almost a  perfect test-bench for reproducing a lot of phenomena from other areas of physics (for review see, e.g. \cite{review-Lewenstein2007-ap}). Particular attention has been  paid to the phenomena from condensed matter and cosmology.  For instance, an advantage to investigate solid-state phenomena like Bloch oscillations \cite{BO-BenDahan1996-prl,BO-Morsch2001-prl}, Landau-Zener tunneling \cite{LZ-Cristiani2002-pra} and Josephson junction \cite{JJ-Gati2006-apb,JJ-Ryu2013-prl} in BEC (instead of proper solid-state structures) lies in the fact that periodic external potential provides an analogue of a perfect crystalline lattice without defects. In the area of cosmology, an interest to both atomic \cite{BEC-Garay2000-prl,BEC-Barcelo2001-cqg,BEC-Liberati2006-cqg,BEC-Recati2009-pra,BEC-Larre2012-pra,BEC-Lahav2010-prl,BEC-Boiron2015-prl} and polaritonic \cite{exc-pol-Solnyshkov2011-prb,exc-pol-Gerace2012-prb,exc-pol-Nguyen2015-prl} BECs comes from the possibility to use it as experimentally attainable system of "analogue gravity" \cite{anal-grav-Barecelo2011-lrr} -- a laboratory model for curved-space quantum theory, e.g., black holes and, in particular, Hawking radiation. 

Hawking radiation represents an additional emission mechanism of particles from a potential well along with classical escaping by external perturbations\cite{Hanggi1990, Rooney2016, Potnis2017} and quantum tunnelling\cite{Caldeira1983, Potnis2017}. A simplified view of this process is that quantum fluctuations create a particle-antiparticle pair nearby the black hole edge \cite{Hawking1974-nature,Hawking1975-cmp}. If one of the pair constituents crosses the event horizon, it never returns, thus giving rise to the emission from the black hole, which in its turn leads to decreasing of black hole energy and mass (for review see, e.g., Ref.\cite{review-Page2005-njp}). However, Hawking emission from nowadays known astronomic black holes is hard to explore because of its weak predicted intensity. Charged black hole amplification of Hawking radiation can take place owing to the resonance in the cavity formed by inner and outer horizons with subsequent black hole lasing  \cite{lasing-Corley1999-prd,lasing-Leonhardt2007}. To realize this idea in BEC a variety of different configurations has been proposed \cite{BEC-Finazzi2010-njp,BEC-Zapata2011-njp,BEC-DeNova2014-njp,BEC-Finazzi2015-prl,BEC-deNova2015-njp},  and later an experimental realization of lasing from BEC was reported \cite{Steinhauer2014-nature}.  It is noteworthy that, through the similarity of wave processes, gravity phenomena can be  also modelled in others physical systems: electromagnetic wave waveguide \cite{elect-wg-Schutzhold2005-prl}, slow light in moving medium \cite{slow-light-Leonhardt200-prl,slow-light-Leonhardt2002-nature}\footnote{Although the possibility to achieve the experimental confirmation of Hawking radiation in this structure is doubtful \cite{slow-light-discussion-Visser2000-prl,slow-light-discussion-Unruh2003-prd}} or  in optical fibers \cite{opt-fibers-theory-Faccio2010-epl,opt-fibers-theory-Philibin2008-science}\footnote{Actually, first experimental evidence of Hawking radiation was observed in this system \cite{opt-fibers-experiments-Belgiorno2010-prl,opt-fibers-experiments-Rubino2011-njp}, although the nature of observed phenomena is under discussion \cite{fiber-optics-discussion-Unruh2012-prd}}, surface waves on moving water \cite{moving-water-Rousseaux2008-njp,moving-water-Weinfurtner2011-prl},  the moving flow with a gradient from subsonic to supersonic flow (so-called acoustic or sonic black hole) \cite{Unruh1981-prl,gas-Auregan2015-prd}. Sonic black holes, after being predicted in 1981, was implemented in quantum liquids like liquid He \cite{He-Jacobson1998-prd}.

Another natural advantage of a BEC system is  that  due to strong two-body interactions BEC matter waves  can be considered as  localized wavepackets or ``solitons'', which can propagate at long distances without losing their shape. Solitons possess a particular interest due to their properties typically associated with particle--like states. Thus, solitons may provide interesting insights about particle-like behavior and particle-antiparticle interaction. For example, an analogue of Hawking radiation in solitons has been studied within the context of a nonlinear Klein-Gordon equation with applications in Josephson junctions \cite{Monica2008-njp}. 
In addition, binary mixtures of Bose-Einstein condensates can support  multicomponent ``vector solitons''\cite{BuschAnglin2001, soliton-review-Brazhnyi2004-mplb,soliton-review-Carretero-Gonzalez2008-nonl,sol-exp-Eiermann2003-prl,sol-exp-Eiermann2004-prl, Cruz2007-pra}.  Loading the BEC in an optical lattice (OL) allows the existence of solitons both in the case of attractive and repulsive interactions \cite{soliton-review-Brazhnyi2004-mplb,soliton-review-Carretero-Gonzalez2008-nonl,sol-exp-Eiermann2003-prl,sol-exp-Eiermann2004-prl}. In such a case,  a  two-component BEC can sustain bright-bright and dark-bright stationary solitons \cite{vector-sol-Ostrovskaya2004-prl}, mixed-symmetry modes and breathers \cite{Cruz2007-pra}. Creation of dark-bright soliton in two-component BEC has been already reported in Ref.\cite{DB-experiment-Becker2008-nat-phys,DB-experiment-Middelkamp2011-pla,DB-experiment-Hamner2011-prl,DB-experiment-Hamner2013-prl}. Meanwhile, the possibility for two-component BEC to be loaded into OL was experimentally demonstrated in Ref.\cite{two-OL-exper-Higbie2005-prl,two-OL-exper-Widera2005-prl}. 

In the present paper, we consider emission of matter wave solitons from a combined parabolic and OL trap \cite{Okulov-X1}. We consider a stationary state of bright-bright solitons in a two-component BEC. Using a periodic sign-varying linear coupling term, we can create a vector soliton, which first and second component chemical potentials are located in the lower and upper edges of the OL spectrum first band.  Under such configuration, the first component will have a positive effective mass, meanwhile the second component will have a negative one. In such sense vector soliton serves as an analogue of a particle-antiparticle pair. When a parabolic trap is loaded, the first component of vector soliton oscillates periodically inside the trap, while the second one accelerates and escapes from it \cite{Okulov-X3}. As an extension of  the Hawking radiation concept, we have called our phenomenon  as  \textit{Hawking-like} emission, in analogy with Ref.\cite{Monica2008-njp}.

 The paper is organized as follows. In section \ref{sec:model} we expose and describe the model of our problem. In section \ref{sec:gap-sol} we represent the stationary state of bright-bright solitons in the two-component BEC and propose a mechanism to create this state. In section \ref{sec:H-emis} we describe the dynamics of two-component soliton in a parabolic trap and the mechanism of a Hawking-like escape of the soliton from the trap.  Conclusions and final remarks are presented in section \ref{conclusions}.

\section{The model and preliminary arguments}
\label{sec:model}

To be specific, we consider a spinor BEC composed of  two hyperfine states, say of
the $|F=1,m_{f}=-1\rangle$ and
$|F=2,m_{f}=1\rangle$ states of  ${}^{87}$Rb
atoms~\cite{Rabi-unitary-Matthews1999-prl,Rabi-unitary-Williams2000-pra} confined at different vertical positions
by transverse parabolic traps and loaded into an optical lattice of cos-like shape. Additionally, a  time-dependent external potential $\gamma(x,t)$ (which is aperiodic in general case) is applied to the condensate. At the same time hyperfine states are coupled by a coordinate- and time-dependent coupling field $\beta(x,t)$, which describes transitions between atomic states. Notice that such coupling can be originated by the external magnetic field.

 We assume a quasi-one-dimensional cigar-shaped condensate.
Then, in the mean-field approximation, the system is described by the Gross-Pitaevskii (GP) equations~\cite{PS}
\begin{eqnarray}
i \frac{\partial \psi_1}{\partial t} =-  \frac{\partial^2 \psi_1}{\partial x^2}-V\cos(2x)\psi_1+\gamma(x,t)\psi_1
+\nonumber\\ \beta(x,t)\psi_2+\left(g_1|\psi_1|^2+ g  |\psi_2|^2 \right) \psi_1,
  \label{eq:tca} \\
i \frac{\partial \psi_2}{\partial t} = -  \frac{\partial^2 \psi_2}{\partial x^2}-V\cos(2x)\psi_2+\gamma(x,t)\psi_2
+\nonumber\\ \beta(x,t)\psi_1+\left(g |\psi_1|^2+g_2|\psi_2|^2\right)\psi_2.
\label{eq:tcb}
\end{eqnarray}
Equations (\ref{eq:tca}) and (\ref{eq:tcb}) are  written
in dimensionless form: OL amplitude $V$, external potential $\gamma(x,t)$ and coupling $\beta(x,t)$ are measured in recoil energy units of $E_R=\hbar^2\pi^2/(2md^2)$ (where $m$ is the atomic mass and $d$ is the OL period), while the coordinate $x$ and time $t$ are measured in units of
$d/\pi$ and $\hbar/E_R$, respectively. At the same time wavefunctions $\psi_j(x)$ are measured in the $a_\bot^2\pi^2/(4d^2|a_{12}|)$ units, where $a_\bot$ is the transverse [in the $(y,z)$-plane] oscillator length, $a_{12}$ is the inter-species s-wave scattering length, thus, $g=\pm 1$ in equations (\ref{eq:tca}) and (\ref{eq:tcb}). Respectively, parameters $g_1=a_1/\left| a_{12}\right|$, $g_2=a_2/\left| a_{12}\right|$ refer to the intraspecies s-wave scattering lengths in the first and second components, measured in the units $|a_{12}|$.

In the absence of additional external force and coupling term, $\gamma(x,t)\equiv 0$, $\beta(x,t)\equiv 0$, periodicity of OL gives rise to band-gap structure of both spectrum $E_n(q)$ and Bloch functions $\varphi_{n,q}(x)$ in the respective linear problem \begin{equation}
E_n(q)\varphi_{n,q}(x)=-d^2\varphi_{n,q}/dx^2-V\cos(2x)\varphi_{n,q}(x) \label{eq:Bloch-lin-problem}
\end{equation} (see figure \ref{fig:state}(a)). Here $n$ is the band number and $q$ is the Bloch wavenumber in the first Brillouin zone, $q\in[-1,1]$. Meanwhile both the spectrum and the Bloch functions are periodic in the reciprocal space with period $2$ (in the chosen units): $E_n(q)=E_n(q+2)$, $\varphi_{n,q}(x)=\varphi_{n,q+2}(x)$. In further considerations we take into account the first band only ($n=1$), whose bottom and top correspond to $E_1(0)$ and $E_1(1)$, respectively, so the index $n$ will be omitted.

\section{Small-amplitude coupled solitons.}
\label{sec:gap-sol}

\subsection{Stationary state}
\begin{figure}
  \begin{center}
   \begin{tabular}{c}
       \includegraphics{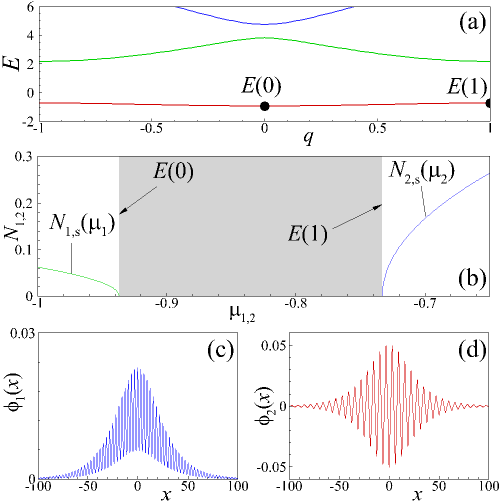}
   \end{tabular}
   \end{center}
\caption{(Color online) (a) Linear band-gap spectrum of OL; (b) First and second component bifurcation diagrams $N_1(\mu_1)$ and $N_2(\mu_2)$, corresponding to the stationary coupled solution (\ref{eq:sts1}), (\ref{eq:sts2}); (c,d) Shapes of coupled solitons (\ref{eq:sts1}), (\ref{eq:sts2}) with chemical potentials $\mu_1=-0.938$, $\mu_2=-0.7317$ (soliton norms $N_1=0.0085$, $N_2=0.0366$, respectively)  -- first $\psi_1(x,0)$ and second $\psi_2(x,0)$ components are depicted in panels (d) and (e), respectively. In all panels the parameters of OL and nonlinearities are: $V=3.0$, $g_1=-2$, $g_2=2$, $g=-1$. The dashed region in the middle of panel (b) are referred to the first band of the OL spectrum.}
\label{fig:state}
\end{figure}

When linear coupling and external force are absent, $\beta\left(x,t\right)=\gamma\left(x,t\right)\equiv0$, the coupled nonlinear Schr\"odinger equations (\ref{eq:tca}) and (\ref{eq:tcb}) possess a family of the stationary solutions $\psi_{j}(x,t)=\phi_{j}(x)\exp(-i\mu_{j}t)$ ($j=1,2$). Here $\mu_{j}$ are the chemical potentials and $\phi_{j}(x)$ are the soliton profiles of the $j$th component, governed by the system of the differential equations
\begin{eqnarray}
\mu_{1}\phi_{1}=-\frac{d^{2}\phi_{1}}{dx^{2}}-V\cos(2x)\phi_{1}+\left(g_{1}\phi_{1}^{2}+g\phi_{2}^{2}\right)\phi_{1},\label{eq:stat-1}		\\
\mu_{2}\phi_{2}=-\frac{d^{2}\phi_{2}}{dx^{2}}-V\cos(2x)\phi_{2}+\left(g\phi_{1}^{2}+g_{2}\phi_{2}^{2}\right)\phi_{2}.\label{eq:stat-2}
\end{eqnarray}		
Without loss of generality, for the solitonic solutions one can consider $\phi_{j}(x)$ to be purely real functions. We also suppose that the chemical potential of the first component soliton $\mu_1$ lies in the semiinfinite gap, $\mu_1<E(0)$, while chemical potential of second component soliton $\mu_2$ is in the first finite gap, $\mu_2>E(1)$. 
When $g_{1}<0$, $g_{2}>0$, equations (\ref{eq:stat-1}) and (\ref{eq:stat-2}) possess a particular approximate solution in the form of low-amplitude \textit{coupled soliton stationary state}. This solution can be obtained \cite{Salerno2002-pra} by the multiply scale expansion method (see supplementary information for the details) and written as
\begin{eqnarray}
\phi_1(x)\approx\left(-\frac{1}{M_0M_1L^2}\frac{M_0g\chi-M_1|g_2|\chi_1}{|g_1g_2|\chi_0\chi_1+\chi^2}\right)^{1/2}\times\nonumber\\
\frac{\varphi_0(x)}{\cosh(x/L)},\label{eq:sts1}\\
\phi_2(x)\approx\left(-\frac{1}{M_0M_1L^2}\frac{M_1g\chi+M_0|g_1|\chi_0}{|g_1g_2|\chi_0\chi_1+\chi^2}\right)^{1/2}\times\nonumber\\
\frac{\varphi_1(x)}{\cosh(x/L)}.\label{eq:sts2}
\end{eqnarray}
Here $M_q = (d^2E(q)/dq^2)^{-1} $ is the effective mass, $\chi_{q}=\int_{-\pi/2}^{\pi/2}|\varphi_{q}(x)|^{4}dx$, $\chi=\int_{-\pi/2}^{\pi/2}|\varphi_{0}(x)\varphi_{1}(x)|^{2}dx$, 
\begin{eqnarray}
L=\{-2M_0[\mu_1-E(0)]\}^{-1/2}=\{-2M_1[\mu_2-E(1)]\}^{-1/2} \label{eq:L-definition}
\end{eqnarray} is the soliton width, which is equal in the first and second component of coupled soliton. 
An example of the coupled soliton shape is represented in figures \ref{fig:state}(c) and (d), showing different symmetries of coupled soliton stationary state components (which are determined by the symmetries of the respective Bloch functions). In particular, while the first component -- the semi-infinite gap soliton -- is  sign-constant, the second component -- the finite gap soliton -- is sign-alternating.

Notice, the coupled soliton chemical potentials $\mu_{1,2}$ are not independent variables: the equal width of solitonic components (\ref{eq:L-definition}) imply the linear character of dependence between chemical potentials, namely
\begin{equation}
\mu_2(\mu_1)=E(1)+\frac{M_0}{M_1}\left[\mu_1-E(0)\right].
\label{eq:mu12}
\end{equation}
Detuning of the first-component chemical potential from the bottom band edge $E(0)>\mu_1$ results into the simultaneous detuning of second-component chemical potential from the top edge of the band $E(1)<\mu_2$ (towards the center of the first gap).

The soliton norm $N_{j,s}=\int\left|\phi_{j}\right|^2dx$ ($j=1,2$) in each component of the stationary state, namely
\begin{eqnarray}
N_{1,s}(\mu_1)=\frac{2^{3/2}}{\pi|M_1|}\left(\frac{E(0)-\mu_1}{M_0}\right)^{1/2}\frac{M_0g\chi-M_1|g_2|\chi_1}{|g_1g_2|\chi_0\chi_1+\chi^2},\\
N_{2,s}(\mu_2)=\frac{2^{3/2}}{\pi M_0}\left(\frac{\mu_2-E(1)}{|M_1|}\right)^{1/2}\frac{M_1g\chi+M_0|g_1|\chi_0}{|g_1g_2|\chi_0\chi_1+\chi^2},
\end{eqnarray}
can be calculated directly from (\ref{eq:sts1}), (\ref{eq:sts2}), substituting rapidly-oscillating Bloch functions $\varphi_{0,1}(x)$ by their average value on the period  $\left\langle\varphi_{0,1}^2(x)\right\rangle=\pi^{-1}$. The results are depicted in figure \ref{fig:state}(b) in the form of bifurcation diagrams, showing the growth of the soliton norm of both components $N_{1,s}$, $N_{2,s}$ with an increase of the detuning of chemical potentials from respective band edges. 

\subsection{Creation of the stationary coupled state.}
\begin{figure}
  \begin{center}
   \begin{tabular}{c}
       \includegraphics{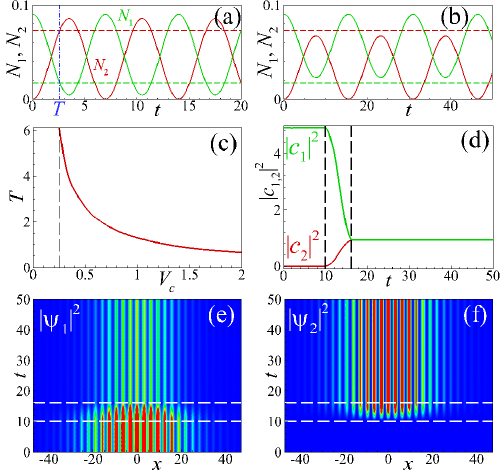}
   \end{tabular}
   \end{center}
\caption{(Color online) (a,b) The temporal dynamics of the number of particles in first (solid green lines) and second  (solid red lines) components for the coupling amplitude $V_c=0.5$ (a) or $V_c=0.2$ (b). The soliton norms in first and second components $N_{1,s}=0.0085$, $N_{2,s}=0.0366$ are depicted by dashed green and red horizontal lines, respectively; (c) exposure time, necessary to achieve state with soliton norms in first and second components $N_{1,s}$, $N_{2,s}$, respectively, versus the coupling amplitude $V_c$; (d--f) temporal evolution of projections $|c_{1,2}(t)|^2$ (d) and spatio-temporal evolution of wavefunctions $|\psi_1(x,t)|^2$ (e) and $|\psi_2(x,t)|^2$ (f) of two-component BEC with linear coupling of the form (\ref{eq:lc-fin}) with $V_c=0.25$, $T_1=10$, $T_2=16.12$ (depicted by two dashed lines in b--d). In all panels other parameters are the same as in figure \ref{fig:state}, and initial stationary state is characterized by the first component chemical potential $\mu_{1,i}=-0.94$ and soliton norm $N_{1,i}=N_{1,s}+N_{2,s}$.}
\label{fig:create}
\end{figure}
Now a natural question arises: how to create the coupled soliton stationary state? In order to answer it, we introduce the concept of an \textit{initial stationary state}: the particular solution of Eqs.(\ref{eq:tca}),(\ref{eq:tcb}) with all atoms concentrated in the first component, 
\begin{eqnarray}
\psi_{1,i}(x,t)=\left(\frac{2[E(0)-\mu_{1,i}]}{|g_1|\chi_0}\right)^{1/2}\times\nonumber\\
\frac{\varphi_0(x)\exp\{-i\mu_{1,i}t\}}{\cosh(\sqrt{-2M_0(\mu_{1,i}-E(0))}x)},\label{eq:stsi1}\\
\psi_{2,i}(x,t)=0.\label{eq:stsi2}
\end{eqnarray}
As the matter of fact, in the initial stationary state, atoms are localized into a bright soliton, whose shape is sign-constant (similar to one, depicted in figure \ref{fig:state}(c)). Its soliton norm can be expressed as
\begin{eqnarray}
N_{1,i}(\mu_{1,i})=\frac{2^{3/2}}{\pi|g_1|\chi_0}\left(\frac{E(0)-\mu_{1,i}}{M_0}\right)^{1/2},\label{eq:init-state1}\\
N_{2,i}=0.\label{eq:init-state2}
\end{eqnarray}
Since the system of equations (\ref{eq:tca}) and (\ref{eq:tcb}) is conservative,  to create a coupled soliton stationary solution with soliton norms $N_{1,s}$ and $N_{2,s}$,  one should start from the initial stationary state (\ref{eq:stsi1}), (\ref{eq:stsi2}) whose soliton norm is equal to
\begin{equation}
N_{1,i}(\mu_{1,i})=N_{1,s}(\mu_{1})+N_{2,s}(\mu_{2}).\label{eq:stsi-dep}
\end{equation}
For the next step, applying nonzero linear coupling between components $\beta(x,t)\ne 0$ for a certain time, one can transfer the atoms from the first component to the second one. Nevertheless, application of a spatially uniform coupling will result in the transferring of the atoms to second component in the same phase, giving rise to similar shapes of solitons in both components. In fact, such resulting soliton will be unstable due to opposite signs of intraspecies nonlinearities $g_1$ and $g_2$. Notwithstanding, the creation of the coupled soliton stationary state with different symmetries in the first and second components (like in figures \ref{fig:state}(c) and \ref{fig:state}(d)) is possible if one uses a spatially-periodic linear coupling with the period equal to $2\pi$,
\begin{equation}
\beta(x,t)=V_c\cos(x).\label{eq:lc}
\end{equation}
Here $V_c$ is the amplitude of coupling. Action of the coordinate-dependent coupling (\ref{eq:lc}) gives rise to the periodical oscillations of the number of particles  $N_1(t)$ and $N_2(t)$ in the first and second components of BEC, as is shown in figures \ref{fig:create}(a) and (b). If one starts with the initial state (\ref{eq:init-state1}), (\ref{eq:init-state2}), a certain amount of particles will be transferred to the second component, and after that will be transferred back, returning to the initial stationary state. By comparing figures \ref{fig:create}(a) and (b) one can see that a growth of the coupling amplitude $V_c$ results in the increase of the oscillation amplitude (larger number of particles is transferred to the second component) and the decrease of the oscillation period.

At certain time moment $T$ after the beginning of oscillations the number of particles in each BEC component will be equal to the number of particles $N_{1,s}$, $N_{2,s}$ (depicted in figures \ref{fig:create}(a) and (b) by horizontal dashed lines) of the coupled soliton stationary state, which we want to create. Furthermore, in the text this time will be referred as the \textit{exposure time}. 
The dependence of the exposure time $T$, which is necessary to achieve the coupled soliton stationary state shown in figures \ref{fig:state}(c) and (d), upon the coupling amplitude $V_c$ is represented in figure \ref{fig:create}(c). As it is evident, an increasing of coupling amplitude results into the decrease of the exposure time. At the same time, there exists a certain amplitude threshold, designated in figure \ref{fig:create}(c) by a vertical dashed line: the desired coupled soliton stationary state can be achieved only for the values of $V_c$ above this threshold.  For the coupling amplitudes below this threshold the number of particles in first and second BEC components never achieve the values $N_{1,s}$, $N_{2,s}$  -- an example of such situation is shown in figure \ref{fig:create}(b).

It is naturally to presuppose that if the coupling is switched off at the time moment, when the number of particles at first and second BEC component are equal to those of the coupled soliton stationary state (i.e. at time moment equal to the exposure time), then the resulted soliton will keep its shape during long time. This fact is confirmed in figures \ref{fig:create}(d)--(f), where the creation of the coupled soliton stationary state is achieved by switching on the coupling at time moment $T_1$ and switching it off after exposure time $T=T_2-T_1$, namely
\begin{equation}
\beta(x,t)=\left\{\begin{array}{c}0,\qquad t<T_1\\V_c\cos(x), \qquad T_1<t<T_2\\0,\qquad t>T_2\end{array}\right.\label{eq:lc-fin}
\end{equation}
The creation of coupled soliton stationary state is evident from figure \ref{fig:create}(d), which represents the projections of the wavefunctions $\psi_j(x,t)$ ($j=1,2$) on the correspondent stationary state $\phi_j(x)$ (see equations (\ref{eq:sts1}),(\ref{eq:sts2})), i.e.
\begin{equation}
c_j(t)=\frac{1}{N_{j,s}}\int_{-\infty}^\infty\psi_j(x,t)\phi_j(x)dx.
\end{equation}
One observes that after coupling is switched off at time moment $T_2$, square modula of these projections are approximately equal to unity during the relatively long integration time. In more details process of the creation of coupled soliton stationary state is depicted in figures \ref{fig:create}(e) and (f). These plots demonstrate both the stability of the initial stationary state at $t<T_1$, and the stability of the created coupled soliton stationary state at $t>T_2$.

\section{Hawking-like emission of matter from the potential well}
\label{sec:H-emis}
\begin{figure}
  \begin{center}
   \begin{tabular}{c}
       \includegraphics{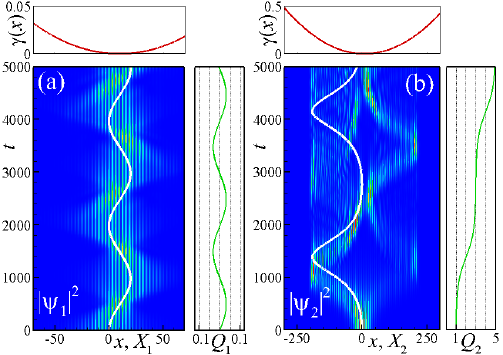}
   \end{tabular}
   \end{center}
\caption{(Color online) Spatio-temporal evolution of particle density $|\psi_1(x,t)|^2$ (a) and $|\psi_2(x,t)|^2$ (b) (depicted by color maps) in two-component BEC  with nonlinearities $g_1=-2$, $g_2=2$, $g=-1$ and loaded into an OL with amplitude $V=3.0$ and parabolic trap (\ref{eq:ex-for}) with $\nu=5\cdot10^{-6}$,  $x_0=10$. The initial condition is the stationary gap soliton with $\mu_1=-0.938$, $\mu_2=-0.7317$, (depicted in figures \ref{fig:state}(c), (d)). Soliton center position in real space $X_1(t)$ and $X_2(t)$ are depicted by white lines in panels (a) and (b), respectively, while soliton center positions in the reciprocal space $Q_1(t)$ and $Q_2(t)$ are represented in lateral figures. Values $X_1(t)$, $X_2(t)$, $Q_1(t)$ and $Q_2(t)$ were obtained from equations (\ref{eq:X1}) -- (\ref{eq:Q2}) with parameters $\omega_0\approx-0.84$, $\omega_1\approx-0.1$, which correspond to OL with amplitude $V=3.0$. The correspondent shape of the parabolic trap $\gamma(x)$ is presented in two upper figures (notice the different spatial scales of the horizontal axes of panels (a) and (b)).}
\label{fig:BO}
\end{figure}
When the additional external potential $\gamma(x,t)$ is applied to the two-component BEC, the dynamics of a localized wavepacket is described  by the semiclassical equations \cite{feng2005introduction}
\begin{eqnarray}
\dot{X_j}=\left.\frac{dE}{dq}\right|_{q=Q_j}, \quad \dot{Q_j}=-\left.\frac{\partial \gamma(x,t)}{\partial x}\right|_{x=X_j}, \label{eq:semiclas}
\end{eqnarray}
where $X_j$ and $Q_j$ denote the center of mass of the matter-wave wavepacket's $j$-component in real and reciprocal space, respectively,
and the overdot stands for the time derivative. Nevertheless, for the validity of the semiclassical equations (\ref{eq:semiclas}) the parameters of the BEC should be  subject to certain limitations. First, the external potential should be \textit{weak}, if compared to the OL: its spatial variation on the period of OL should be much smaller than the half-width of the energy band, i.e. $|\gamma(x+\pi,t)-\gamma(x+\pi,t)|\ll|E(1)-E(0)|/2$. In the opposite case the motion of the wavepacket will be suppressed due to appearance of the Wannier-Stark ladder \cite{feng2005introduction}. This limitation is applicable both in linear and nonlinear cases. Second limitation is specific for the nonlinear case only: the soliton width $L$ (see equation (\ref{eq:L-definition})) should be much larger, than the OL period, $L\gg\pi$. Taking into account the inverse dependence between the amplitude of the soliton and its width (see equations (\ref{eq:sts1}) and (\ref{eq:sts2})), this requirement can be reformulated as the necessity to use the low-amplitude solitons only. When both these conditions are satisfied, the semiclassical equations (\ref{eq:semiclas}) describe well \cite{BOnonl-Salerno2008-prl,BOnonl-Bludov2009-epl} the dynamics of the soliton in BEC with OL, but only when the modulational instability condition is met. In the modulationally stable regions soliton is destroyed, hence the semiclassical equations are not valid any more.

When the parabolic time-independent potential
\begin{equation}
\gamma(x,t)=\nu(x-x_0)^2
\label{eq:ex-for}
\end{equation}
is applied to the two-component soliton, equations (\ref{eq:semiclas}) can be written as
\begin{eqnarray}
X_j=x_0-\frac{\dot{Q_j}}{2\nu}, \label{eq:semiclas-2a}\\
\ddot{Q_j}+2\nu\omega_1\pi\sin\left(\pi Q\right)=0. \label{eq:semiclas-2b}
\end{eqnarray}
In equations (\ref{eq:semiclas-2a}), (\ref{eq:semiclas-2b}) we approximated the band structure $E(q)$ leaving only two leading terms of Fourier expansion, i.e.
\begin{equation}
E(q)=\omega_0+\omega_1\cos(\pi q).\label{eq:band-appr}
\end{equation}
Equation (\ref{eq:semiclas-2b}) is the differential equation, which describes the oscillation of simple pendulum. Thus, under initial conditions $X_1(0)=X_2(0)=0$, $Q_1(0)=0$, $Q_2(0)=1$, system of equations (\ref{eq:semiclas-2a}), (\ref{eq:semiclas-2b}) possesses an \textit{exact} solution
\begin{eqnarray}
X_1(t)=x_0\left[1-{\rm cn}\left(\frac{\pi\nu x_0}{k_1}t,k_1\right)\right],\label{eq:X1}\\
Q_1(t)=\frac{2}{\pi}\arcsin\left[k_1\,{\rm sn}\left(\frac{\pi\nu x_0}{k_1}t,k_1\right)\right],\label{eq:Q1}\\
X_2(t)=x_0\left[1-\frac{1}{{\rm dn}\left(\frac{\pi\sqrt{2\nu \omega_1}}{k_2}t,k_2\right)}\right],\label{eq:X2}\\
Q_2(t)=\frac{2}{\pi}{\rm am}\left(\frac{\pi\sqrt{2\nu \omega_1}}{k_2}t,k_2\right).\label{eq:Q2}
\end{eqnarray}
In the above equations
$$
k_1=\left(\frac{\nu x_0^2}{2\omega_1}\right)^{1/2}, \qquad k_2=\left(1+\frac{\nu x_0^2}{2\omega_1}\right)^{-1/2},
$$
are elliptic moduli, ${\rm cn}\left(t,k\right)$, ${\rm dn}\left(t,k\right)$ are the Jacobi elliptic functions, ${\rm am}\left(t,k\right)$ is the Jacobi amplitude.

As it follows from equations (\ref{eq:Q1}) and (\ref{eq:Q2}), in the reciprocal space, the first component soliton center should exhibit periodical oscillations in the vicinity of $Q_1=0$ (see the lateral panel of figure \ref{fig:BO}(a)). At the same time the soliton center in the second component (the lateral panel of figure \ref{fig:BO}(b)) is a increasing function of time\footnote{with the initial condition $Q_2(0)=1$, equation (\ref{eq:semiclas-2b}) resembles the rotating pendulum}. In coordinate space soliton centers of both first and second components exhibit periodic oscillations (see Eqs.(\ref{eq:X1}) and (\ref{eq:X2})). It is worth noting first component of soliton oscillates in the vicinity of the parabolic trap minimum $x_0$, with oscillation period $\tau_1=4K(k_1)(2\pi^2\nu\omega_1)^{-1/2}$ and amplitude $\xi_1=x_0$ (white lines in figure \ref{fig:BO}(a)), the solitonic second component moves in the region $x<0$ with  period $\tau_2=2k_2K(k_2)(2\pi^2\nu\omega_1)^{-1/2}$ and amplitude $\xi_2=(\sqrt{(2\omega_1+\nu x_0^2)/\nu}-x_0)/2$. Again, it is depicted in figure \ref{fig:BO}(b) by white lines.

These predictions are confirmed by direct numerical integration of equations (\ref{eq:tca}), (\ref{eq:tcb}).  Figure \ref{fig:BO} shows the resulting spatio-temporal evolution of two-component soliton. Nevertheless, while periodical oscillations of solitonic first component are stable, i.e., soliton keeps its shape during long time of evolution, the second component soliton is destroyed  after certain time, less than one oscillation period $\tau_2$ (see figures \ref{fig:BO}(a) and (b), respectively). This  phenomenon can be explained, if to compare the numerical results with the semiclassical ones. The first component soliton oscillates in the narrow interval in reciprocal space $\sim-0.03\le Q_1 \le \sim 0.03$ (lateral panel in figure \ref{fig:BO}(a)), inside which effective mass is always positive, $M_{Q_1}>0$. As a result, the modulational instability condition $M_{Q_1}g_1<0$ is kept at every moment of time, preventing the soliton from the destruction -- the semiclassical treatment in the first component is valid in every stage of the process. In contrast, the second component $Q_2$ (lateral panel in figure \ref{fig:BO}(b)) passes through all values inside the first Brilloin zone, even where the instability condition $M_{Q_2}g_2<0$ is not met, resulting in the destruction of soliton (notice the above-mentioned non-validity of solutions (\ref{eq:X1})--(\ref{eq:Q2}) after the soliton destruction in spite of their formal existence). The quantitative comparison of the numerical results with semiclassical ones shows good correspondence between predicted amplitudes of oscillation in coordinate space and reasonable correspondence between oscillation periods. The discrepancy between periods of oscillation takes place both due to the inexactness of the band approximation (\ref{eq:band-appr}) and due to inexactness of semiclassical equations (\ref{eq:semiclas}) in the nonlinear case.

At the initial stage of evolution $t\gtrsim 0$ the positions of the soliton in reciprocal space $Q_1$ and $Q_2$ increase due to negative $d\gamma(x)/dx$ at $x=0$ (as it  follows from equation (\ref{eq:semiclas})). In its turn positive $dE(q)/dq$  at $q=0+0$ causes first-component soliton motion to the $x$-positive direction, while negative derivative $dE(q)/dq$ at $q=1+0$ causes that second component moves to the $x$-negative direction, i.e. outwards the parabolic trap center. Similar result for the one-component soliton was demonstrated in Ref.\cite{Sakaguchi2005-mcs}. If the second component of soliton is accelerated during the initial stage, and at certain coordinate $x_{EH}$ the action of external trap is switched off, i.e., $d\gamma(x)/dx=0$, then it will continue its motion with constant velocity, escaping from the parabolic trap. This coordinate $x_{EH}$ can be considered as analogue of event horizon in black hole. In general, the coordinate $x_{EH}$ should be less than the position of second-component soliton center in real space at quarter-period $X_2(\tau_2/4)$. This requirement comes from the necessity to stop action of external force, when the second-component soliton center in the reciprocal space is inside the interval $1<Q_2<1.5$ (where effective mass is negative) in order to prevent soliton from further destruction. 

\begin{figure}
  \begin{center}
   \begin{tabular}{c}
       \includegraphics{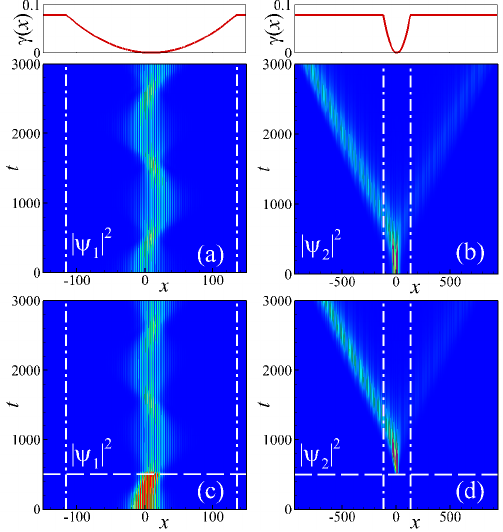}
   \end{tabular}
   \end{center}
\caption{(Color online) Spatio-temporal evolution of particle density $|\psi_1(x,t)|^2$ ((a) and (c)) and $|\psi_2(x,t)|^2$ ((b) and (d)), obtained from the numerical integration of equations (\ref{eq:tca}),(\ref{eq:tcb}) with finite-width parabolic trap (\ref{eq:ex-for-fin}) and with the linear coupling term $\beta(x,t)=0$ ((a) and (b)), or of the form (\ref{eq:lc-fin}) ((c) and (d)) with $V_c=0.25$, $T_1=500$, $T_2=506.12$ (exposure time is the same as in figure \ref{fig:create}). The initial conditions in (a) and (b) correspond to those in figure \ref{fig:BO}, and in panels (c) and (d) are the same as in figure \ref{fig:create}. In all panels the parameters of OL, finite-width parabolic trap, and nonlinearities are $V=3.0$, $\nu=5\cdot10^{-6}$, $x_0=10$, $D=80\pi$, $g_1=-2$, $g_2=2$, $g=-1$. In all panels edges of the finite-width parabolic trap are depicted by vertical dash-and-dotted lines (the correspondent shape of the finite-width parabolic trap $\gamma(x)$ is presented in two upper figures). In panels (c) and (d) time moments $T_1$ and $T_2$ are depicted by horizontal dashed lines (indistinguishable in the scale of panels). }
\label{fig:smooth}
\end{figure}

As an example, this scenario can be realised, when the soliton is placed inside the finite-width parabolic potential (compare with equation (\ref{eq:ex-for}))
\begin{eqnarray}
\gamma(x,t)=\left\{\begin{array}{c} \nu(x-x_0)^2,\qquad |x-x_0|<D/2,\\ \nu(D/2-x_0)^2,\qquad |x-x_0|\ge D/2\end{array}\right.,
\label{eq:ex-for-fin}
\end{eqnarray}
where $D$ is the width of the potential well. The shape of the potential (\ref{eq:ex-for-fin}) is depicted in the upper panels of figure \ref{fig:smooth}. In the frame of the above-mentioned formalism, edges of the finite-width potential (\ref{eq:ex-for-fin}), $x_{EH1}=x_0-D/2$ and $x_{EH2}=x_0+D/2$, can be considered as event horizon analogues. As seen from figures \ref{fig:smooth}(a) and (b), first-component soliton oscillates periodically (see Fig.\ref{fig:smooth}(a)) inside the finite-width potential well (\ref{eq:ex-for-fin}). Meanwhile, second-component soliton is accelerated during its movement in the negative direction of $x$-axis (see figure \ref{fig:smooth}(b)), and after crossing the potential-well edge $x_{EH1}$, depicted by vertical dash-and-dotted line, continues its motion with constant velocity.
%

Moreover, we can start from the situation when all the BEC atoms are initially concentrated in the first component (like in equations (\ref{eq:stsi1}),(\ref{eq:stsi2})), and then apply, for a finite time, the spatially-periodic linear coupling, which will transfer a portion of atoms to the second component. In other words, we use the same method, as described in section \ref{sec:gap-sol}, but apply it not to the stationary soliton, but to the oscillating soliton  inside the finite-width potential well (see figure \ref{fig:smooth}(c)). In this case atoms, transferred to the second component, will constitute the gap soliton with negative effective mass, which in its turn will escape from the finite-width potential well, as demonstrated in figure \ref{fig:smooth}(d).

\section{Conclusions}
\label{conclusions}
We described a mechanism of stimulated emission of matter waves, in form of bright solitons, from the two-component BEC, loaded into the OL, which is combined with an external parabolic potential. The similarity (in general terms) between the Hawking emission from the black hole and the soliton escape from the parabolic trap is defined by the fact, that we use the bright-bright soliton. Chemical potentials of first and second BEC component lie nearby the opposite edges of the first band of OL spectrum. As a consequence, signs of the effective masses are also opposite, and such type of low-amplitude soliton can be considered as an analogue of a particle-antiparticle pair in Hawking emission. We demonstrated, that this low-amplitude bright-bright two-component soliton can be created by partial transferring of atoms from one to another BEC component, using spatially-periodic linear coupling term, whose period equals to the double OL period. Being loaded into the finite-width parabolic trap, one component of such bright-bright soliton with positive effective mass exerts periodic oscillations nearby the trap center, while another component, with negative effective mass, is gradually accelerated and moves in the direction of parabolic trap growth. If soliton with negative effective mass passes the finite-width parabolic trap edge, it escapes from the trap and never returns to the initial point. 
The model proposed above has certain differences with Hawking radiation. Firstly,  our analogue of a particle-antiparticle pair, the vector soliton, is created by transferring atoms from first to second components of BEC. In conventional Hawking emission, by contrast, the particle-antiparticle pair is formed from vacuum due to quantum fluctuations. Secondly, in our phenomenon the event horizon is defined somehow artificially as a tunable edge of  a potential well.

\section*{Acknowledgment}
Y.V.B. acknowledges the support from Portuguese Foundation for Science and Technology (FCT) through Grant No. UID/FIS/04650/2013. M.A.G.-N. thanks for the financial support of FONDECYT project 11130450.

\bibliographystyle{unsrt}
\bibliography{haw_emis_bib}

\end{document}